\documentclass{aa}    

\usepackage[colorlinks=true, urlcolor=blue, citecolor=blue, linkcolor=blue]{hyperref}
\usepackage{caption}
\usepackage{graphicx}
\usepackage{txfonts}
\usepackage{lipsum}
\usepackage{subcaption} 

\usepackage{lscape}             
\usepackage{placeins}           
\usepackage[svgnames]{xcolor}  

\begin{document}

\title{Changing-look active galactic nuclei from SDSS, LAMOST, and DESI surveys}

\author{
Guohai Chen\inst{1,2,3,4,5},
Wenxin Yang\inst{1,2,4,5}, 
Xuhong Ye\inst{1,4,5},
Zhiqiang Chen\inst{6},
Zhiyuan Pei\inst{1,4,5},
Hubing Xiao\inst{7}
\\ and Junhui Fan\inst{1,3,4,5}\thanks{E-mail: fjh@gzhu.edu.cn (J.H. Fan)
}}

\institute{Center for Astrophysics, Guangzhou University, Guangzhou 510006, China
\and Dipartimento di Fisica e Astronomia “G. Galilei”, Università di Padova, Via F. Marzolo, 8, I-35131 Padova, Italy
\and Abastumani Observatory, Mount Kanobili, 0301 Abastumani, Georgia
\and Greater Bay Brand Center of the National Astronomical Data Center, Guangzhou 510006, China
\and Astronomy Science and Technology Research Laboratory of Department of Education of Guangdong Province, Guangzhou 510006, China
\and School of Astronomy and Space Science, Nanjing University, Nanjing 210023, China
\and Key Lab for Astrophysics, Shanghai Normal University, Shanghai 200234, China
\\}

\date{Received September 30, 20XX}

\abstract
{Although more than 1000 optical changing-look active galactic nuclei (CLAGNs) have been reported to date, their physical origin remains unclear, and known repeating CLAGNs (RCLAGNs) are still rare. Expanding the CLAGN sample, especially RCLAGNs, is therefore important for constraining their underlying mechanism.}
{This work aims to systematically identify new CLAGNs and investigate the occurrence of repeated changing-look (CL) behaviour by combining multi-epoch spectroscopy from the Sloan Digital Sky Survey (SDSS), the Large Sky Area Multi-Object Fiber Spectroscopic Telescope (LAMOST), and the Dark Energy Spectroscopic Instrument (DESI). We further examined the accretion-state evolution and transition timescales of the identified sources to constrain the mechanisms responsible for CL activity.}
{We cross-matched spectroscopic observations from SDSS and LAMOST and searched for significant changes in the broad emission lines between different epochs. DESI spectra were used as a third spectroscopic epoch to search for repeated CL transitions. We also examined the evolution of the sources in the $\log M_{\rm BH}$--$\log (L_{\rm bol}/L_{\rm Edd})$ plane and estimated rest-frame upper limits on the transition timescales from the intervals between spectroscopic observations.}
{We identify 45 CLAGNs, of which 40 are newly reported. The sample is dominated by turn-off events, comprising 43 turn-off and 2 turn-on sources. This may be because Type 2 AGNs either lack a detectable broad-line region or have their broad emission lines obscured by circumnuclear dust, making turn-on events more difficult to identify. Using DESI spectra as a third spectroscopic epoch, we identified 12 RCLAGNs. Of the 14 newly identified CLAGNs with three-epoch spectroscopic coverage, 7 objects, i.e. 50\%, are confirmed as RCLAGNs. In the previously reported sample, 5 out of 16 objects with available DESI spectra, corresponding to $\sim$31\%, exhibit repeating CL behaviour. The relatively high incidence of repeated CL behaviour suggests that CL transitions are associated with recurrent physical processes, such as accretion-rate fluctuations or accretion-disk instabilities. In the $\log M_{\rm BH}$--$\log (L_{\rm bol}/L_{\rm Edd})$ plane, the RCLAGNs display a clear high--low--high accretion-state evolution, indicating a close connection between repeated CL behaviour and recurrent variations in accretion power. The rest-frame upper limits on the transition timescales are $\sim$10 yr for the first transition and 4 yr for the second transition.}
{Our results expand the known CLAGN and RCLAGN samples and demonstrate that repeated spectral-type transitions may be common and detectable when sufficient spectroscopic epochs are available. The observed accretion-state evolution supports intrinsic changes in the accretion flow as an important driver of CL activity. The different upper limits obtained for the first and second transitions reflect the different temporal baselines of the spectroscopic surveys rather than intrinsically different physical transition timescales.}


\titlerunning{CLAGNs from SDSS, LAMOST, and DESI surveys}
\authorrunning{Guohai Chen et al.}
\maketitle
\nolinenumbers


\section{Introduction}

Active galactic nuclei (AGNs) are powered by supermassive black holes at their centres
\citep{Kormendy2013ARA&A..51..511K}. From the accretion of surrounding gas and dust, an accretion disk forms, in which a portion of the gravitational potential energy of the infalling material is converted into electromagnetic radiation. The accretion disk is surrounded by the broad-line region (BLR), and then by the narrow-line region (NLR) on larger scales. A torus-like dusty structure may enshroud the system on intermediate scales \citep{Urry1995PASP..107..803U}.
AGNs are observationally divided into two main types based on optical spectroscopic features. Type 1 AGNs display both broad emission lines (BELs; line widths of 1000–20000 km s$^{-1}$) and narrow emission lines (NELs; 300–1000 km s$^{-1}$), while Type 2 AGNs show only NELs. According to the unified model, this difference arises from orientation: in Type 2 AGNs, the BELs are thought to be obscured from our line of sight by a dusty torus
\citep{Antonucci1993ARA&A..31..473A, Riffel2006A&A...457...61R}.
However, it is now established that a population of `true' Type 2 AGNs exists; this population appears to lack a BLR rather than being obscured Type 1 AGNs
\citep{Tran2001ApJ...554L..19T, Wang2007ApJ...660.1072W, Elitzur2009ApJ...701L..91E, Shen2014Natur.513..210S, Suzy2022agn..book....1C}.
In recent years, spectroscopic observations have revealed a class of optical changing-look AGNs (CLAGNs) in which the BELs in the optical spectra can appear (turn on) or disappear (turn off) on timescales ranging from months to years, as reviewed by \cite{Ricci2023NatAs...7.1282R} and other works
\citep{Yang2018ApJ...862..109Y, Trakhtenbrot2019ApJ...883...94T, Zeltyn2022ApJ...939L..16Z}. Since their discovery, this phenomenon has attracted significant attention in astronomy.

A robust identification of CLAGNs requires both the emergence or disappearance of BELs and significant photometric variability. To date, more than 1000 optical CLAGNs have been identified, primarily through two main approaches \citep{Tohline1976ApJ...210L.117T, Penston1984MNRAS.211P..33P, Cohen1986ApJ...311..135C, Storchi1993ApJ...410L..11S, Aretxaga1999ApJ...519L.123A, Eracleous2001ApJ...554..240E, Denney2014ApJ...796..134D, Yang2018ApJ...862..109Y, Trakhtenbrot2019ApJ...883...94T, Potts2021A&A...650A..33P, Zeltyn2022ApJ...939L..16Z, Jin2022ApJ...926..184J, Yang2023ApJ...953...61Y, Guo2024ApJS..270...26G, Guo2025ApJS..278...28G, Dong2025ApJ...986..160D, Chen2026ApJS..282...28C}.
One approach relies on repeat spectroscopic observations from multi-epoch spectral surveys, such as the Sloan Digital Sky Survey \citep[SDSS;][]{Paris2018A&A...613A..51P}, the Large Sky Area Multi-Object Fiber Spectroscopic Telescope \citep[LAMOST;][]{Cui2012RAA....12.1197C}, the Dark Energy Spectroscopic Instrument \citep[DESI;][]{DESI22026AJ....171..285D}, and the Six-degree Field Galaxy Survey \citep[6dFGS;][]{Jones2009MNRAS.399..683J}. Sources exhibiting changes in BELs are identified, and photometric variability is further examined to rule out BEL variations caused by instrumental effects or other observational uncertainties.
The other approach begins with the selection of candidates based on large-amplitude photometric variability, which are subsequently confirmed as CLAGNs through follow-up spectroscopic observations that capture such BEL transitions \citep[e.g.][]{Yang2018ApJ...862..109Y, Yang2025ApJ...980...91Y, MacLeod2019ApJ...874....8M, Sheng2020ApJ...889...46S, Zhu2024MNRAS.530.3538Z, Zhu2025MNRAS.536.2715Z}. In both approaches, visual inspection is often applied as a final verification step.

Studies of CLAGNs indicate that there are three main mechanisms for their changing-look  (CL) behaviour: intrinsic variations in the accretion rate, dust obscuration effects, and tidal disruption events (TDEs). 
Each mechanism has observational limitations and cannot explain all cases. The accretion-rate variations are the most widely supported mechanism, with the accretion disk state transitions driving the observed spectral changes
\citep{Noda2018MNRAS.480.3898N, Ruan2019ApJ...883...76R, Sniegowska2020A&A...641A.167S, Feng2021ApJ...916...61F, Ricci2023NatAs...7.1282R, Wu2023ApJ...958..146W}.
In contrast, the obscuration scenario attributes BEL variations to dusty material moving into and out of the line of sight \citep{Holt1980ApJ...241L..13H}. For most optical CLAGNs, the transition timescales are much shorter than the expected dynamical timescales of obscuring dust \citep{LaMassa2015ApJ...800..144L, Runnoe2016MNRAS.455.1691R, Sheng2017ApJ...846L...7S}, and this scenario cannot explain the significant variability observed in their light curves \citep{MacLeod2016MNRAS.457..389M, Ruan2016ApJ...826..188R, Runnoe2016MNRAS.455.1691R}. Nevertheless, \cite{Zeltyn2022ApJ...939L..16Z} reported a rapid CLAGN transition on a rest-frame timescale of $\lesssim 2$ months whose spectral evolution favoured obscuration, although the short timescale and mid-infrared variability made this interpretation physically challenging.
The TDE scenario \citep{Merloni2015MNRAS.452...69M, Padmanabhan2021A&A...656A..47P} can explain only a limited number of cases
\citep{Merloni2015MNRAS.452...69M} and has difficulty accounting for repeating CLAGNs (RCLAGNs).

It remains unclear whether all CLAGNs undergo repeated CL transitions. Despite the growing number of identified CLAGNs, known recurrent CL events are still rare because their confirmation requires multi-epoch spectroscopic evidence of multiple state changes, which is often obtained through long-term monitoring over years to decades. The nearby Seyfert galaxy Mrk 1018 provides one of the earliest and most striking examples of observed repeating CL behaviour. Its H$\beta$ emission increased by more than a factor of six within a few years, before fading again and disappearing several decades later
\citep{Cohen1986ApJ...311..135C, McElroy2016A&A...593L...8M}. Long-term spectroscopic monitoring further revealed that Mrk 1018 experienced a complete sequence of spectral-type transitions, evolving through Seyfert types 1.0, 1.2, 1.5, 1.8, and 2.0 over a span of approximately 45 years \citep{Lu2025ApJS..276...51L}. A similar evolutionary pattern has been observed in Mrk 590. The H$\beta$ line first emerged in 1989, vanished by 2003 \citep{Denney2014ApJ...796..134D}, and subsequently reappeared in 2017 together with strong continuum variability \citep{Raimundo2019MNRAS.486..123R, Mandal2021MNRAS.508.5296M}. More recently, \cite{Palit2025MNRAS.540L..14P} reported that Mrk 590 is once again transitioning towards a type 1.2 state, closely resembling its spectral classification in the 1990s. To date, the number of confirmed RCLAGNs remains small, with only about 50 sources reported in the literature \citep{Wang2024ApJ...966..128W, Wang2025ApJ...981..129W, Lyu2025A&A...693A.173L, Dong2026ApJ..1006..133D, Dong2025ApJ...986..160D}.

Therefore, identifying more RCLAGNs through long-term photometric monitoring and multi-epoch spectroscopic observations is essential for constraining the physical origin and characteristic timescales of the CL phenomenon, and constitutes one of the main motivations of this work. Low-redshift ($z<0.9$) CLAGNs have been systematically identified using SDSS in combination with DESI \citep{Guo2024ApJS..270...26G, Guo2025ApJS..278...28G}, as well as SDSS together with LAMOST \citep{Dong2025ApJ...986..160D}. \cite{Chen2026ApJS..282...28C} explored CLAGNs at higher redshifts using SDSS and DESI; however, a systematic search for high-redshift CLAGNs based on SDSS and LAMOST remains lacking.
To address this gap, we conducted a systematic search for CLAGNs using SDSS and LAMOST without relying on [O III]-based flux recalibration, and further incorporated DESI spectra to investigate repeating CL behaviour and construct an enlarged RCLAGN sample.

The paper is organized as follows. Section \ref{Data} presents the spectroscopic and photometric data, as well as the spectral fitting and flux calibration methods, adopted in this work. Section \ref{Sample Selection} describes the procedure for selecting CLAGNs and RCLAGNs. Sections \ref{DISCUSSION} and \ref{Summary} provide the discussion and summary, respectively. We adopted a flat $\Lambda$ cold dark matter cosmology with 
$H_{0}=71~{\rm km~s^{-1}~Mpc^{-1}}$ and 
$\Omega_{\rm m}=0.27$ \citep{Komatsu2011ApJS..192...18K}.

\section{Data and analysis} \label{Data}
\subsection{Spectroscopic data}

The SDSS spectroscopic observations were primarily carried out with the Sloan Foundation 2.5 m telescope \citep{Gunn2006AJ....131.2332G} at Apache Point Observatory in New Mexico, USA, as part of several successive phases of the survey.
We studied spectra from the SDSS Seventeenth Data Release \citep[SDSS DR17;][]{Abdurro2022ApJS..259...35A}, which contains a total of 5801200 spectroscopic observations. The SDSS spectroscopic pipeline classifies objects as galaxies (`GALAXY'), quasars (`QSO'), or stars (`STAR') through template fitting to observed spectra \citep{Hutchinson2016AJ....152..205H, Bolton2012AJ....144..144B}. Of these, 1192886 spectra are classified as stars and 4608314 as galaxies or quasars.
The SDSS spectra were obtained over multiple survey phases using different instruments. The spectra from SDSS-I and SDSS-II cover a wavelength range of 3800–9100 \AA, while those from SDSS-III, obtained mainly by the Baryon Oscillation Spectroscopic Survey (BOSS), extend the coverage to 3600–10400 \AA, with a typical spectral resolution of 
$R \sim 2000$ \citep{Adelman2008ApJS..175..297A, Alam2017MNRAS.470.2617A}.

LAMOST is a 4 m reflecting Schmidt telescope located at Xinglong Observatory, China, designed for large-scale spectroscopic surveys. It is equipped with 4000 fibres over a $5^\circ$ field of view, enabling the simultaneous acquisition of thousands of spectra \citep{Cui2012RAA....12.1197C, Zhao2012RAA....12..723Z}.
Our analysis is based on the LAMOST Twelfth Data Release (LAMOST DR12), which contains a total of 12605485 spectra. Of them, 12236412 are classified as stars, and 369073 as galaxies or quasars by the LAMOST pipeline.
We used only the low-resolution LAMOST spectra, which span the optical wavelength range from 3700 to 9000 \AA\, and have a characteristic spectral resolution of $R \sim 1800$. The LAMOST spectrograph adopts a dual-arm design, with a blue arm covering 3700–5900 \AA\, and a red arm covering 5700–9000 \AA\,
\citep{Luo2012RAA....12.1243L, Du2016ApJS..227...27D}.

DESI is a Stage IV ground-based spectroscopic survey carried out with the 4 m Mayall Telescope at Kitt Peak National Observatory \citep{Levi2013arXiv1308.0847L, DESI2016arXiv161100036D, DESI2016arXiv161100037D}. The instrument is equipped with 5000 robotic fibres over an 8 deg$^{2}$ field of view, allowing for the simultaneous observation of thousands of targets. DESI spectra span a wavelength range of 3600–9800 \AA, divided into three channels: blue (3600–5900 \AA), green (5660–7220 \AA), and red (7470–9800 \AA), with typical spectral resolutions of $R \sim 2100$, 3200, and 4100, respectively \citep{DESI2016arXiv161100037D, DESI2022AJ....164..207D}.
We used DESI spectra from Survey Validation and early Year 1 observations released as part of DESI Data Release 1 (DR1). We further adopted the AGN or QSO Value-Added Catalog (VAC) for DESI DR1, which provides improved galaxy and quasar classifications and redshift estimates based on the Redrock pipeline, QuasarNet, and FastSpecFit measurements \citep{Alexander2023AJ....165..124A, Chaussidon2023ApJ...944..107C, Moustakas2023ascl.soft08005M}.

\subsection{Cross-matching of SDSS and LAMOST data}
\label{matching}
We performed a positional cross-matching between the LAMOST DR12 and SDSS DR17 catalogues using the TOPCAT software \citep{Taylor2005ASPC..347...29T}. A matching radius of $2^{\prime\prime}$ was adopted, and all possible matches within this radius were retained. We further required a redshift consistency of $\Delta z < 0.002$ between the SDSS and LAMOST spectra. Applying these criteria, we obtained a total of 272924 matched SDSS–LAMOST spectral pairs. From this parent sample, we retained only sources classified as GALAXY or QSO, which yielded 25904 GALAXY–QSO and QSO–QSO spectral pairs for subsequent analysis. These spectral pairs form the parent sample for the subsequent CLAGN selection procedure described in Sect. \ref{Sample Selection}.

\subsection{Initial flux recalibration of LAMOST spectra}
\label{Recalibration of LAMOST spectra}

The response curves of the LAMOST spectrographs have been removed during the data reduction process, but the resulting spectra are not absolutely flux calibrated \citep{Du2016ApJS..227...27D}. We followed \cite{Wang2018MNRAS.474.1873W} and \cite{Liu2022ApJ...927...57L} to recalibrate the spectral fluxes by comparing synthetic photometry derived from the spectra with SDSS photometric measurements.
Specifically, each LAMOST spectrum is convolved with the SDSS $g$, $r$, and $i$ filter response curves \citep{Fukugita1996AJ....111.1748F} to obtain synthetic magnitudes, which are then compared with the corresponding SDSS \texttt{psfMag}. A zeroth- or first-order polynomial was adopted depending on the presence or absence of a wavelength-dependent trend in the magnitude differences between the $g$, $r$, and $i$ bands, and the resulting correction was applied to the original LAMOST spectrum to obtain the $gri$-based flux-calibrated spectrum. Therefore, we required all 25904 spectral pairs to have available SDSS $g$, $r$, and $i$ \texttt{psfMag} measurements. This selection left 19704 spectral pairs.

However, since AGNs are variable, the use of SDSS photometry to recalibrate LAMOST spectra obtained at different epochs may introduce additional uncertainties. To further assess the robustness of this method, we adopted an independent flux correction approach based on the narrow [O III] $\lambda5007$ emission line, which is widely used as a relative flux calibrator under the assumption that narrow-line emission remains constant over the relevant timescales \citep[e.g.][]{Guo2024ApJS..270...26G, Guo2025ApJS..278...28G, Dong2025ApJ...986..160D}.
For this comparison, we restricted the analysis to LAMOST spectra with redshifts $z<0.9$, where the [O III] $\lambda5007$ line is covered by both the LAMOST and SDSS spectra. Each LAMOST spectrum was then rescaled to match the [O III] $\lambda5007$ flux measured from its corresponding SDSS spectrum.

\begin{figure*}
\centering
\includegraphics[width=0.8\textwidth]{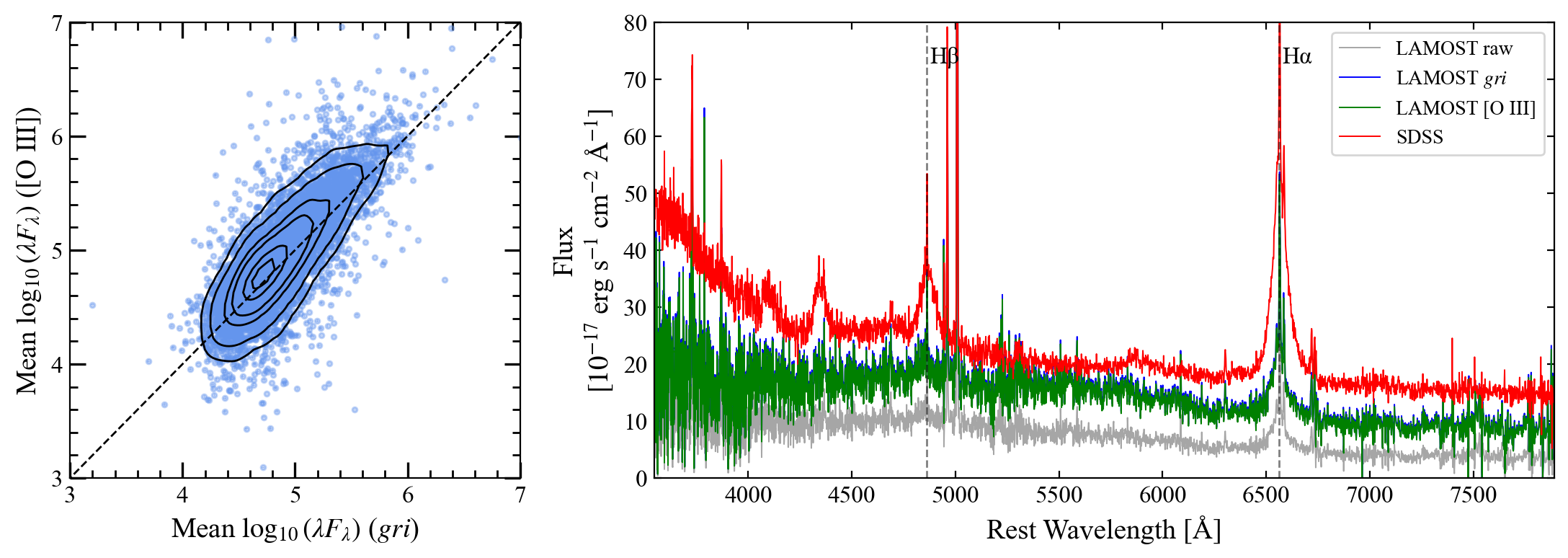}
\caption{\textit{Left}: Comparison of the mean log$_{10}(\lambda F_{\lambda})$ values derived from the $gri$-based and [O III]-based flux recalibration methods for LAMOST spectra. The blue points show individual sources with uncertainties, and the black contours represent the kernel density estimation. The dashed line indicates the one-to-one relation.
\textit{Right}: Example spectrum of J103228.85+350207.0. The grey line shows the original LAMOST spectrum, and the blue and green lines represent the $gri$-based and [O III]-based recalibrated LAMOST spectra, respectively. The SDSS spectrum is shown in red for comparison. Vertical dashed lines mark the positions of prominent emission lines.}
    \label{fig.1}
\end{figure*}

A comparison of the two recalibration methods is shown in Fig. \ref{fig.1} with J103228.85+350207.0 as an example (right panel): the $gri$-based recalibrated LAMOST spectrum (blue) nearly overlaps with the [O III]-based recalibrated spectrum (green), whereas the original spectrum (grey) differs substantially. This shows that the two independent approaches can produce comparable flux recalibrations for individual sources.

A statistical comparison is presented in the left panel of Fig. \ref{fig.1}. For each source, we computed the mean value of $\log_{10}(\lambda F_{\lambda})$ over the common wavelength range after applying the $gri$-based and [O III]-based corrections, respectively. The best-fitting linear relation has a slope of $0.97 \pm 0.01$ and an intercept of $0.18 \pm 0.06$, with a Pearson correlation coefficient of $r=0.68$ and a $p$-value of $<10^{-6}$, but also with a large intrinsic scatter of $\sim0.38$ dex. The scatter indicates that the SDSS $gri$ \texttt{psfMag}-based recalibration is more suitable for preliminary sample screening than for final flux recalibration. 
Since the spectra span a long time baseline, obtaining quasi-simultaneous photometry requires searches across multiple optical databases, some of which do not support efficient bulk queries. We therefore adopted the $gri$-based recalibration for the initial screening of the full sample and applied a more accurate correction using quasi-simultaneous photometric data only to the selected candidates (see Sect. \ref{Intercalibration}).

\begin{figure*}
\centering
\includegraphics[width=0.6\textwidth]{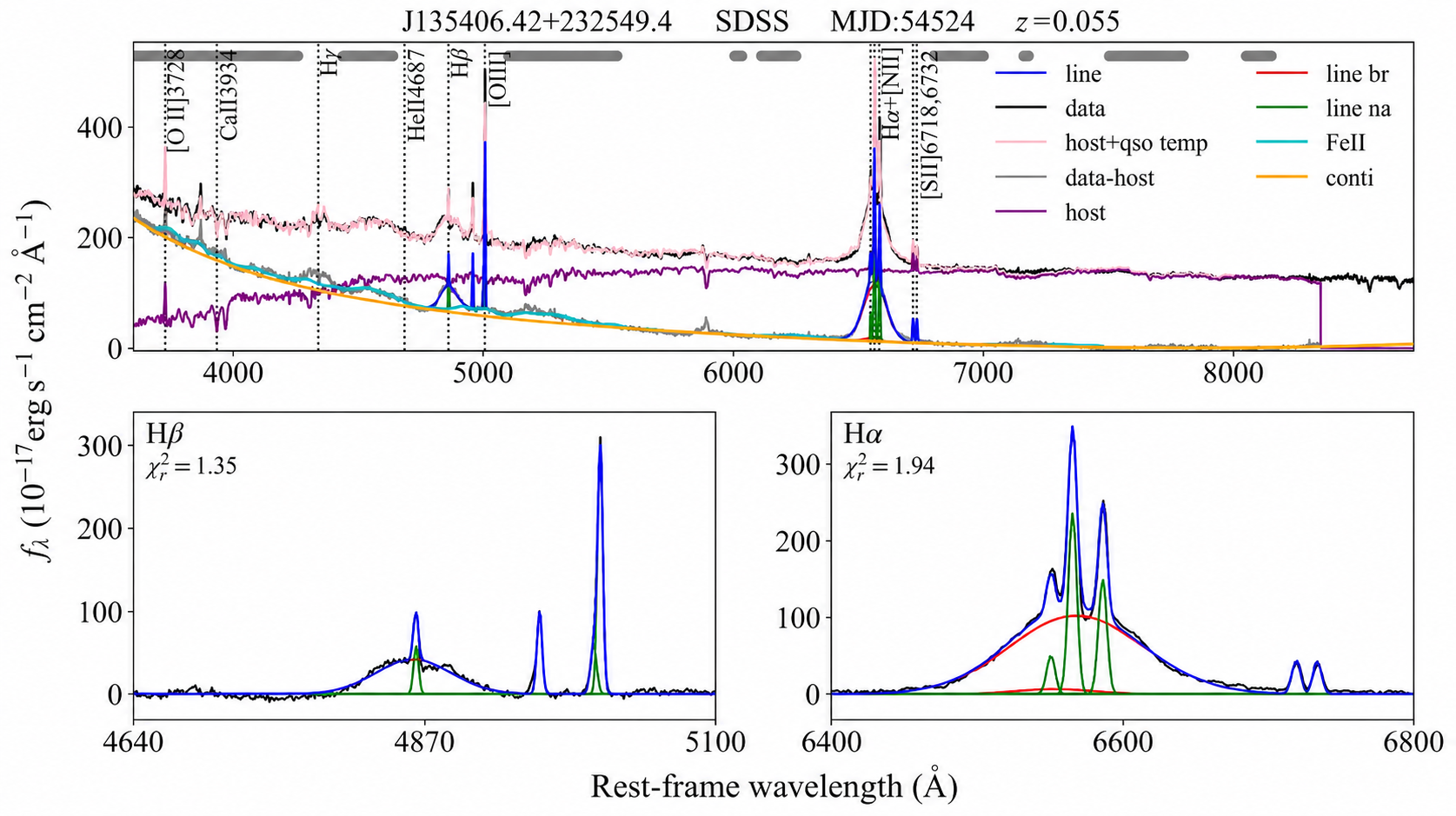}
\label{fig.2a}
\includegraphics[width=0.6\textwidth]{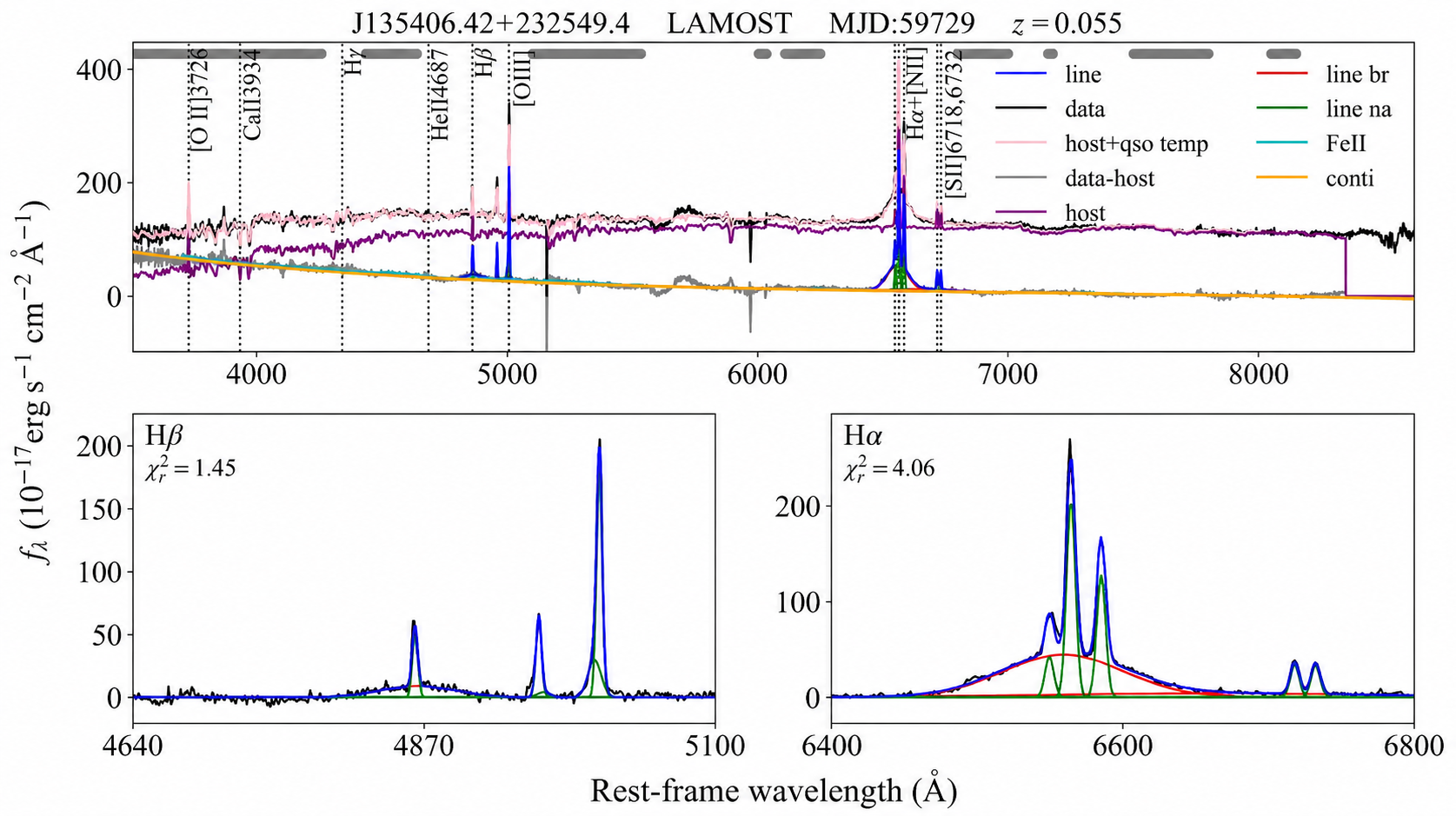}
\label{fig.2b}
\caption{
Spectral fitting results for J135406.42+232549.4 from two different epochs.
\textit{Top}: SDSS spectrum in the high state, showing prominent broad H$\beta$ and H$\alpha$ emission lines.
\textit{Bottom}: LAMOST spectrum in the low state, where the broad H$\beta$ component has disappeared and the broad H$\alpha$ emission is significantly weakened.
The observed spectra are plotted in black. The continuum model is shown in orange, the BEL and NEL components in red and green, respectively, and the combined best-fit model in blue. Grey horizontal segments mark the wavelength regions used for continuum fitting.}
    \label{fig.2}
\end{figure*}

\subsection{Spectral fitting}

We used QSOFITMORE \citep{Fu2021zndo...5810042F} to fit the optical spectra in this work. This code performs spectral decomposition by combining continuum models, Fe II templates, and emission-line components, and determines the best-fit parameters through $\chi^2$ test. QSOFITMORE is developed from PyQSOFit \citep{Guo2018ascl.soft09008G}, but it is more flexible and can process optical spectra from different instruments without requiring the SDSS data format. In our recent works \citep{Chen2024ApJS..271...20C, Chen2025MNRAS.544.1926C, Chen2026A&A...707A.347C}, we applied this tool to analyse publicly available optical spectra of Fermi blazars. The key parameters and a brief overview of the spectral fitting procedure are presented below. Detailed descriptions can be found in \cite{Guo2018ascl.soft09008G}, \cite{Shen2019ApJS..241...34S} and \cite{Fu2021zndo...5810042F}.

 The target spectra are first corrected for Galactic extinction using the extinction curve of \citet{Cardelli1989ApJ...345..245C} and the dust map of \citet{Schlegel1998ApJ...500..525S}. To reduce host-galaxy contamination, the stellar component is removed via principal component analysis following \citet{Yip2004AJ....128..585Y, Yip2004AJ....128.2603Y}. The continuum is then modelled with four components: a power law, a third-order polynomial, and optical and ultraviolet Fe II emission templates \citep{Boroson1992ApJS...80..109B, Vestergaard2001ApJS..134....1V, Shen2019ApJS..241...34S}. After subtracting the best-fit continuum, the emission lines are fitted within predefined wavelength windows. The H$\alpha$ and H$\beta$ regions are fitted over 6400–6800 \AA\, and 4640–5100 \AA, respectively, with their broad components modelled by three Gaussians, while the narrow components and associated forbidden lines (e.g. [O III], [N II], and [S II]) are fitted with single Gaussians. The Mg II and C IV lines are fitted over 2700–2900 \AA\, and 1500–1700 \AA, respectively; the Mg II broad component is modelled with two Gaussians and its narrow component with one Gaussian, while the C IV broad component is modelled with three Gaussians \citep{Shen2019ApJS..241...34S}.

Figure \ref{fig.2} shows the spectral fitting results for the source J135406.42+232549.4 based on two independent observations. The decomposition reveals clear differences between the two epochs. The SDSS spectrum corresponds to a high state, where strong broad components of both H$\beta$ and H$\alpha$ are detected. In contrast, the LAMOST observation represents a low state: the broad H$\beta$ emission has nearly disappeared, and the broad H$\alpha$ component is substantially reduced in strength.

\subsection{Photometric data}
\label{Intercalibration}
Although the spectra show clear BEL variations, establishing a CLAGN typically requires corresponding photometric variability. We therefore examined multi-epoch light curves to confirm that the observed changes are intrinsic rather than caused by observational effects. In the optical band, we utilized data from the Catalina Real-Time Transient Survey \citep[CRTS;][]{Drake2009ApJ...696..870D}, the All-Sky Automated Survey for SuperNovae \citep[ASAS-SN;][]{Kochanek2017PASP..129j4502K}, and the \textit{Zwicky} Transient Facility \citep[ZTF;][]{Bellm2019PASP..131a8002B}. CRTS has provided V-band monitoring since 2007, ASAS-SN has conducted all-sky observations since 2013, and ZTF has delivered high-cadence $g$- and $r$-band data since 2018. In the mid-infrared, we used data from the Wide-field Infrared Survey Explorer \citep[WISE;][]{Wright2010AJ....140.1868W}, including both the original WISE mission (since 2010) and its reactivation phase (NEOWISE; since 2013).

For the CRTS data, we applied a density-based filtering procedure to remove outliers. Specifically, we constructed a magnitude histogram with 20 bins, identified the most populated bin corresponding to the dominant magnitude distribution, and retained only data points within $\pm$0.5 mag of the peak-bin centre.
For ZTF, we first selected measurements with \texttt{catflags} = 0, indicating good photometric quality. The data were then grouped by filter ($g, r, i$), and outliers were removed in each band using the same density-based method as adopted for CRTS.
For ASAS-SN, we retained only measurements with \texttt{Quality} = 'G' and excluded upper-limit detections. We further removed measurements with unusually large photometric uncertainties using a 3$\sigma$ criterion applied to the distribution of magnitude errors.
For WISE/NEOWISE, we retained only measurements with reliable photometric quality flags (\texttt{ph\_qual} = AA, AB, BA, or BB).

\subsection{Photometric intercalibration and final spectral flux correction}

Given the differences in photometric apertures and zero-points between the surveys, intercalibration was required before the light curves could be combined. We adopted the ZTF $r$-band light curve as the reference system and aligned the CRTS and ASAS-SN measurements to it.
Since the variability amplitude differences between nearby optical bands are small \citep{Wilhite2005ApJ...633..638W}, we assumed that the intrinsic variability amplitudes are comparable and corrected only for a constant magnitude offset between surveys \citep{Dong2025ApJ...986..160D}.
We applied a simple additive scaling \citep{Peterson1991ApJ...368..119P},
$m^{\prime} = m + C$,
where $m$ and $m^{\prime}$ are the magnitudes before and after alignment, and $C$ is a constant offset relative to the ZTF $r$-band system.
To determine the calibration offsets, we adopted a sequential two-step procedure.
First, the ASAS-SN V-band light curve was aligned to the ZTF $r$-band reference system. We constructed matched data pairs by selecting measurements obtained within $\pm$5 days. The magnitude differences of all matched pairs are calculated, and the mean difference is adopted as the offset between ASAS-SN V and ZTF $r$, denoted as $C_{V-r}$.
Second, the CRTS V-band light curve is aligned to the ASAS-SN V-band light curve using the same $\pm$5 day matching criterion. The mean magnitude difference of all matched pairs is taken as the offset between CRTS and ASAS-SN, denoted as $C_{V-V}$.

The SDSS \texttt{psfMag}-based recalibration described in Sect. \ref{Recalibration of LAMOST spectra} is used only in the initial automatic screening stage, where it provides a uniform flux reference for identifying spectral pairs with significant BEL variations. For the sources that remain after visual inspection, we further refined the LAMOST flux calibration using quasi-simultaneous optical photometry. Specifically, for each visually confirmed candidate, we matched the LAMOST spectroscopic epoch with the nearest ZTF $r$-band observation and rescaled the LAMOST spectrum to ensure consistency with the corresponding photometric flux level. This two-step strategy enables efficient candidate selection from the full SDSS--LAMOST matched sample while reducing systematic uncertainties in the LAMOST flux scale and ensuring a more reliable confirmation of CLAGN behaviour.

\begin{figure*}
\centering
\includegraphics[width=0.45\textwidth]{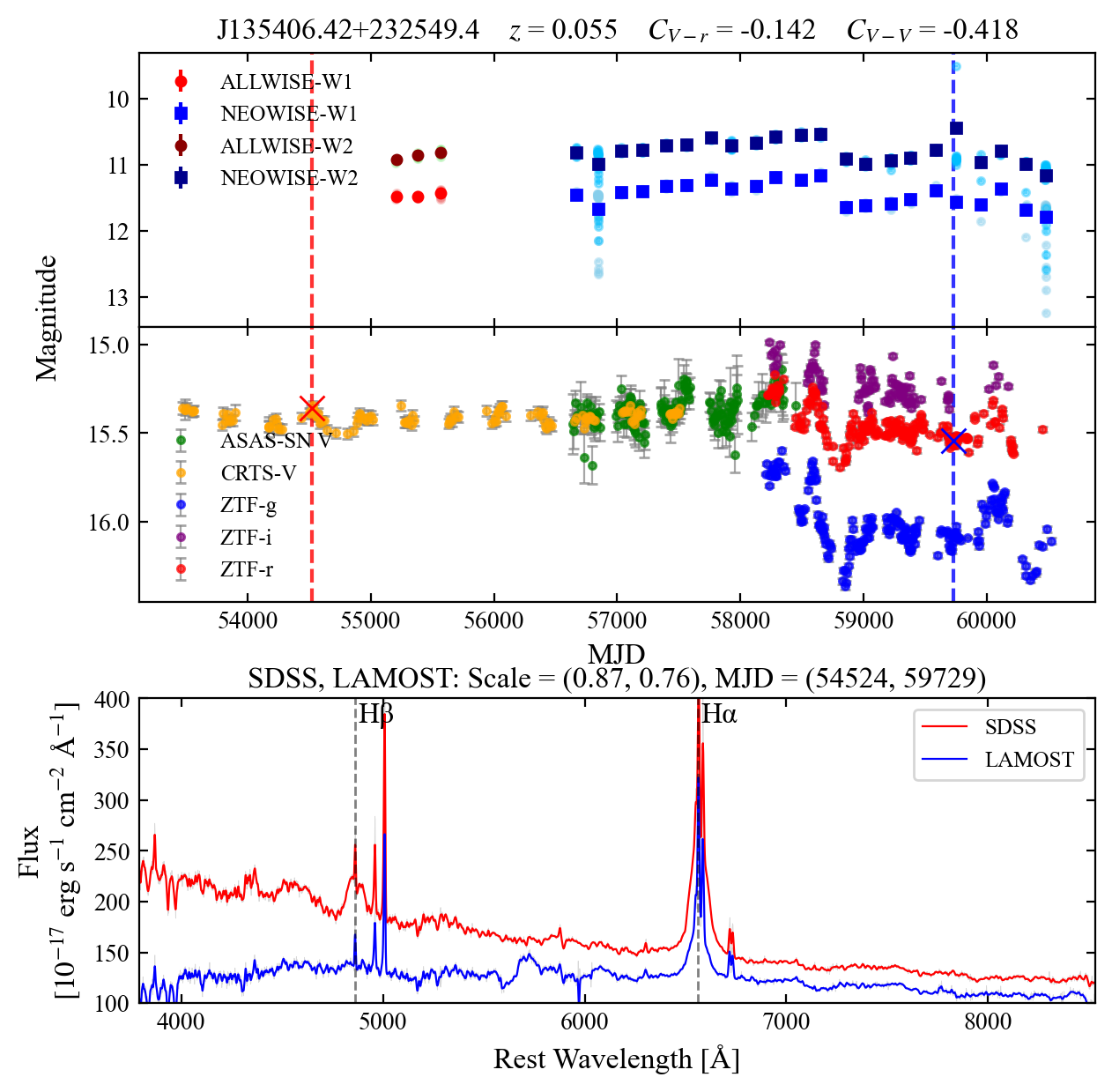}
\includegraphics[width=0.45\textwidth]{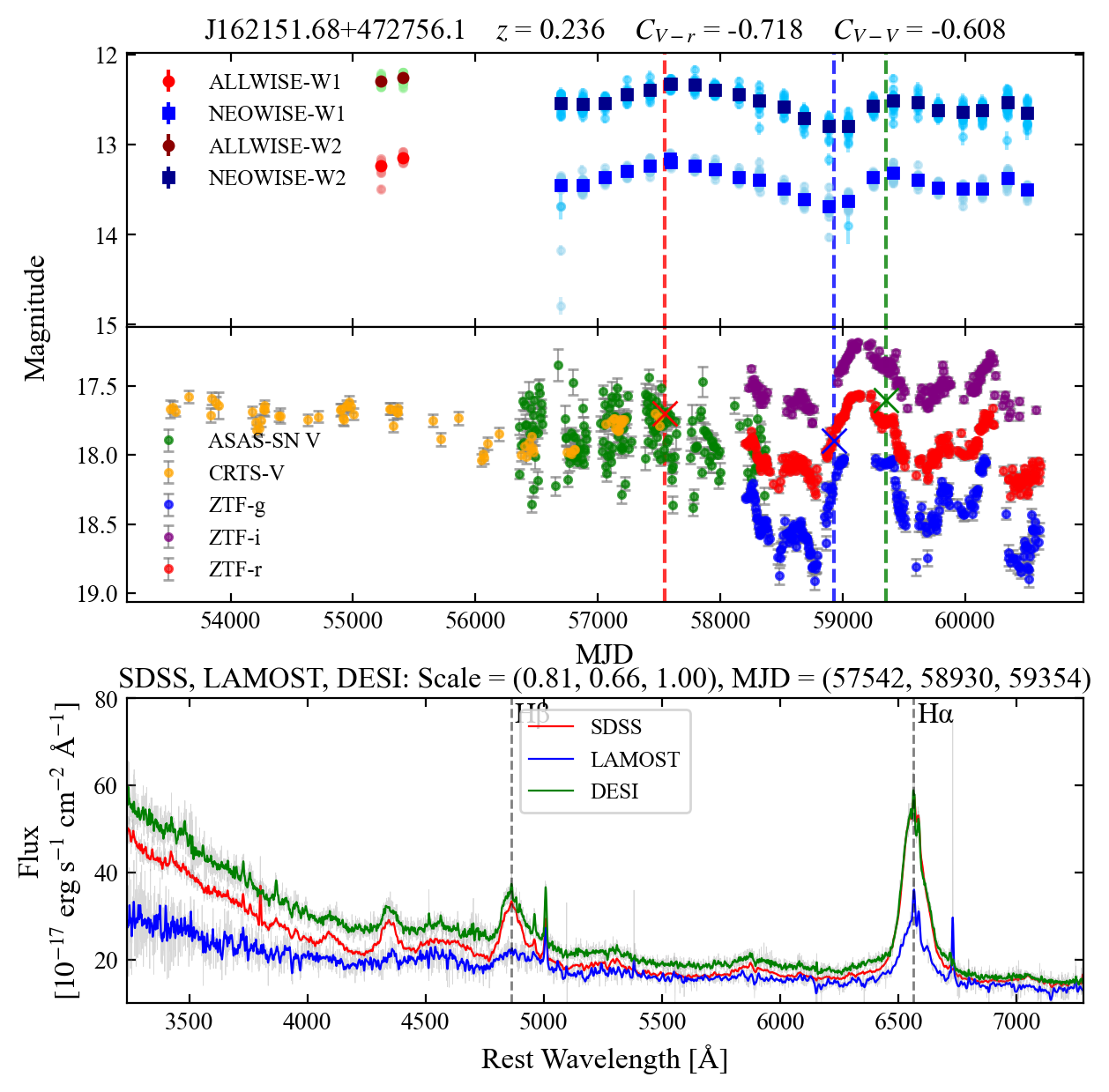}
\caption{
Multiwavelength variability and spectral comparisons of two CLAGNs.
\textit{Panel (a)}: CLAGN J135406.42+232549.4, with SDSS and LAMOST spectroscopic observations.\textit{ Panel (b)}: RCLAGN J162151.68+472756.1, for which SDSS, LAMOST, and DESI spectra are available.
In each panel, the top subplot presents mid-infrared light curves from ALLWISE (W1 red; W2 dark red) and NEOWISE (W1 blue; W2 dark blue). Faint points show individual measurements, and solid symbols represent binned data.
The middle subplot shows inter-calibrated optical light curves from ZTF ($g$ blue; $r$ red; $i$ purple), ASAS-SN (V green), and CRTS (V orange). ASAS-SN V is shifted to the ZTF $r$-band, and CRTS V is aligned to ASAS-SN V. Applied offsets are indicated in the titles, where $C_{V-r}$ denotes the constant magnitude offset used to align ASAS-SN V to the ZTF $r$-band system, and $C_{V-V}$ denotes that used to align CRTS V to ASAS-SN V. Vertical dashed lines mark the epochs of SDSS (red), LAMOST (blue), and, where available, DESI (green) spectroscopy.
The bottom subplots compare optical spectra obtained from SDSS, LAMOST, and, where available, DESI. Original spectra are shown in light grey, with Gaussian-smoothed spectra overplotted in colour. The SDSS and LAMOST spectra are scaled to the ZTF $r$-band reference system, while DESI spectra throughout this work are shown without additional flux rescaling.
The complete figure set, comprising 45 images, is available as described in the data availability section.}
\label{fig.3}
\end{figure*}

Examples are shown in Fig. \ref{fig.3}, which presents the multiwavelength light curves and multi-epoch optical spectra of two representative sources. The left panel shows the CLAGN J135406.42+232549.4, for which the SDSS spectrum obtained during a high state exhibits BELs, whereas the LAMOST spectrum taken during a low state shows the disappearance of the broad components. The right panel presents the RCLAGN J162151.68+472756.1, which is covered by SDSS, LAMOST, and DESI spectroscopy and displays a recurrent change in the BEL. The spectral scaling factors and the photometric intercalibration constants, $C_{V-r}$ and $C_{V-V}$, are labelled in the figure. The complete figure set, comprising the corresponding plots for all 45 sources, is available on Zenodo.

\begin{table*}
\fontsize{9.5}{12}\selectfont
\setlength{\tabcolsep}{5pt} 
\centering
\caption{Procedure for selecting CLAGNs from the SDSS and LAMOST surveys.}
\label{Table1}
\begin{tabular}{llc}
\hline
\hline
Step & Selection Criteria & Number \\
\hline
1. SDSS--LAMOST cross-matching
    & Angular separation $< 2''$ and redshift difference $\Delta z < 0.002$
    & 272924 \\

2. Spectral-type pairs
    & Classification: GALAXY--QSO or QSO--QSO
    & 25904 \\

3. Photometric quality
    & Valid SDSS $gri$ \texttt{psfMag}
    & 19704 \\

4. Broad-line variability selection
    &  Peak-flux variation $R_{\rm p} > 0.4$ and integrated-flux variation $R_{\rm s} > 0.4$
    & 9469 \\

5. Visual inspection
    & Appearance or disappearance of broad emission lines
    & 45\\
\hline
\end{tabular}
\tablefoot{This table summarizes the step-by-step selection of CLAGNs from the SDSS and LAMOST spectroscopic surveys. $R_{\rm p}$ and $R_{\rm s}$ denote the fractional variations in the peak flux and integrated flux of the BEL, respectively. The final SDSS--LAMOST CLAGN sample consists of 45 objects, of which 40 are newly identified in this work.}
\end{table*}

\section{Sample selection and identification}
\label{Sample Selection}

\subsection{Selection of CLAGNs}

For each spectral pair, we retained it if at least one of the two spectra exhibits a reliable BEL component. In this subsection we take the H$\beta$ line as an example; the same criteria were applied to the other BELs.
We relied on the spectral fitting results provided by QSOFITMORE. However, automated fitting produces unphysical results, such as extremely broad components or weak broad-line contributions dominated by noise. To ensure robustness, we applied the following criteria on the broad H$\beta$ component, following \cite{Dong2025ApJ...986..160D} with minor modifications:

1. Objects with best-fit broad H$\beta$  full width at half maximum (FWHM) exceeding 13000 km s$^{-1}$ are excluded to avoid spurious fitting solutions.

2. The relative strength between the narrow and broad components must satisfy $F_{\mathrm{H}\beta,\mathrm{na}}/{F_{\mathrm{H}\beta,\mathrm{br}}} < 5$, where $F_{\mathrm{H}\beta,\mathrm{na}}$ and ${F_{\mathrm{H}\beta,\mathrm{br}}}$ denote the peak flux densities of the narrow and broad H$\beta$ components, respectively.

3. The broad H$\beta$ component must contribute significantly to the total flux around the line centre, requiring $(F_{\mathrm{H}\beta,\mathrm{br}} + F_{\mathrm{cont}})/{F_{\mathrm{cont}}} > 1.2$, where $F_{\mathrm{cont}}$ is the continuum flux density at the central wavelength of H$\beta$.

To quantify the variability of the H$\beta$ broad line, we evaluated both the peak-flux and integrated-flux differences between two spectral epochs. The fractional variation in peak flux is defined as

\begin{equation}
R_p =
\frac{
\left| F_{\mathrm{H}\beta,\mathrm{br}}^{\rm A} -
F_{\mathrm{H}\beta,\mathrm{br}}^{\rm B} \right|
}{
\max \left(
F_{\mathrm{H}\beta,\mathrm{br}}^{\rm A},
F_{\mathrm{H}\beta,\mathrm{br}}^{\rm B}
\right)
}.
\end{equation}

\noindent Here,
$F_{\mathrm{H}\beta,\mathrm{br}}^{\rm A}$ and
$F_{\mathrm{H}\beta,\mathrm{br}}^{\rm B}$
represent the peak fluxes of the broad H$\beta$ component in the two spectra of each pair.

Similarly, the fractional variation in integrated flux is defined as

\begin{equation}
R_s =
\frac{
\left| S_{\mathrm{H}\beta,\mathrm{br}}^{\rm A} -
S_{\mathrm{H}\beta,\mathrm{br}}^{\rm B} \right|
}{
\max \left(
S_{\mathrm{H}\beta,\mathrm{br}}^{\rm A},
S_{\mathrm{H}\beta,\mathrm{br}}^{\rm B}
\right)
},
\end{equation}

\noindent where
$S_{\mathrm{H}\beta,\mathrm{br}}^{\rm A}$ and
$S_{\mathrm{H}\beta,\mathrm{br}}^{\rm B}$
denote the corresponding integrated fluxes.

Applying the above broad-line reliability criteria together with the variability requirements of $R_p>0.4$ and $R_s>0.4$ \citep{Dong2025ApJ...986..160D}, we selected 9469 spectral pairs that exhibit significant BEL variations. This automatic selection corresponds to Step 4 in Table \ref{Table1}. We then visually inspected these spectral pairs to identify sources with clear BEL changes, requiring the appearance or disappearance of broad H$\beta$ or broad H$\alpha$ emission lines, together with corresponding brightening or dimming in the optical and infrared light curves. For all visually confirmed sources with significant BEL variations, we recalibrated the LAMOST spectra using the quasi-simultaneous photometric data introduced in Sect. \ref{Intercalibration}. 

This final verification yields a sample of 45 CLAGNs, corresponding to Step 5 in Table \ref{Table1}. We further compared these sources with previously reported CLAGN samples \citep[e.g.][]{Ruan2016ApJ...826..188R, MacLeod2016MNRAS.457..389M, MacLeod2019ApJ...874....8M, Yang2018ApJ...862..109Y, Sheng2020ApJ...889...46S, Zeltyn2024ApJ...966...85Z, Guo2024ApJS..270...26G, Guo2025ApJS..278...28G, Dong2025ApJ...986..160D, Chen2026ApJS..282...28C} and confirmed that 40 of them have not been reported as CLAGNs before. An example of a newly identified CLAGN is shown in the left panel of Fig. \ref{fig.3}.

\subsection{Identification of RCLAGNs}

To search for repeating CL behaviour, we first focused on the 40 newly identified CLAGNs in this work and cross-matched them with DESI DR1 using the same positional and redshift criteria adopted in Sect. \ref{matching}, requiring an angular separation smaller than $2^{\prime\prime}$ and a redshift difference of $\Delta z < 0.002$. Of these 40 sources, 14 have available DESI spectra and therefore provide three-epoch spectroscopic coverage from SDSS, LAMOST, and DESI.

For these 14 sources, we examined BEL variability across the three spectroscopic epochs using the same criteria adopted for the CLAGN selection. We required significant broad-line variability, with $R_p > 0.4$ and $R_s > 0.4$, together with visual confirmation of the appearance or disappearance of BEL components. A source was classified as an RCLAGN if it exhibits two distinct spectral-state transitions across the three epochs, corresponding to either an `appearance--disappearance--reappearance' or a `disappearance--appearance--disappearance' sequence of BELs. Of the 14 newly identified CLAGNs with DESI spectroscopy, 7 satisfy these criteria and are classified as RCLAGNs. This corresponds to an RCLAGN detection rate of 50\% among the newly identified CLAGNs with DESI spectroscopy.

Our original aim was to report RCLAGNs among the newly identified CLAGNs in this work. However, the high fraction of RCLAGNs motivated us to apply the same procedure to the 51 SDSS--LAMOST CLAGNs reported by \citet{Dong2025ApJ...986..160D}. Sixteen of these sources have available DESI spectra, five of which show clear repeated CL transitions. Although these five objects were previously known as CLAGNs, they exhibit repeated CL transitions and are therefore included in the final RCLAGN sample. Combining the two subsamples, we obtained 12 RCLAGNs. An example is shown in the right panel of Fig. \ref{fig.3}.

In terms of photometric variability, all RCLAGNs show significant brightness changes consistent with their spectral evolution. Multiwavelength light curves from ZTF, ASAS-SN, CRTS, and WISE/NEOWISE reveal pronounced brightening or dimming during the corresponding epochs, with variability trends that closely track the strengthening or weakening of the BELs. This consistency further rules out instrumental or observational effects and supports an intrinsic origin of the observed changes.

\subsection{The final CLAGN and RCLAGN samples}

The final sample reported in this work consists of 45 CLAGNs identified from the SDSS--LAMOST matched spectra, including 40 newly reported sources, together with 12 RCLAGNs confirmed through the incorporation of DESI spectroscopy. The basic properties of these objects are summarized in Table \ref{Table2}, including their coordinates, redshifts, spectroscopic observation epochs, black hole masses, Eddington ratios, variable emission lines, and spectral-transition classifications.

\begin{table*}
\begin{center}
\fontsize{8}{11}\selectfont
\setlength{\tabcolsep}{2pt} 
\caption{Properties of the 33 CLAGNs and 12 RCLAGNs (extract).}
\label{Table2}
\begin{tabular}{ccccccccccccc}
\hline
\hline
Name    &       RA    &       Dec    &       $z$     &       MJDs    &       log$M_{\rm BH, 1}$ &       log$M_{\rm BH, 2}$      &       log$\lambda_{\rm Edd, 1}$       &       log$\lambda_{\rm Edd, 2}$        &       log$\lambda_{\rm Edd, 3}$       &       Line    &       Trans. & Ref.  \\

(1)&(2)&(3)&(4)&(5)&(6)&(7)&(8)&(9)&(10)&(11)&(12)&(13)\\
\hline

J000934.85+180343.0     &       2.39525 &       18.06195        &       0.309   &       56199-56947     &       $8.34  \pm 0.04$      &               &       $-0.56  \pm 0.06 $      &       $-1.12  \pm 0.07 $     &               &       H$\beta$        &       on      &       TW      \\
J001700.15+113720.8     &       4.25063 &       11.62246        &       0.072   &       58507-59144     &       $7.27  \pm 0.03$      &               &       $-0.99  \pm 0.05 $      &       $-1.26  \pm 0.03 $     &               &       H$\beta$        &       off     &       TW      \\
J013944.73+243121.6     &       24.93638        &       24.52267        &       0.162   &       57360-58397     &       $7.64  \pm 0.05$      &               &       $-1.16  \pm 0.08 $      &       $-1.28  \pm 0.06 $     &               &       H$\beta$        &       off     &       TW      \\

... & ... & ....& ... & ... & ....& ... & ... & .... & ... & .... & ... \\
\hline
\end{tabular}
\end{center}
\tablefoot{
(1) LAMOST name;
(2) right ascension (RA);
(3) declination (Dec);
(4) redshift ($z$);
(5) the Modified Julian Dates of the available spectroscopic observations, listed chronologically as MJD$_1$--MJD$_2$--MJD$_3$ and corresponding to the epochs in columns (8)--(10);
(6) and (7) the logarithm of the black hole mass in units of solar mass, estimated from the broad H$\alpha$ emission line in the first and second high spectral states, respectively, when available;
(8)--(10) the logarithm of the Eddington ratios corresponding to the first, second, and third spectroscopic epochs, respectively;
(11) the BELs that show significant variations across multiple spectroscopic observations;
(12) the first changing state identified between the first and second spectroscopic epochs;
(13) reference for the source identification, where `TW' denotes this work and `D25' denotes \citet{Dong2025ApJ...986..160D}. The full table is available at the CDS.}
\end{table*}

\section{Results and discussion}
\label{DISCUSSION}

\subsection{Comparison with previous CLAGN searches}

\citet{Dong2025ApJ...986..160D} identified 51 CLAGNs by cross-matching SDSS and LAMOST, with the redshift limited to $z < 0.9$ due to the requirement of using the [O\,III] $\lambda5007$ emission line for flux recalibration of LAMOST spectra. In this work, we adopted an alternative calibration strategy based on SDSS $gri$ photometric magnitudes \citep{ Fukugita1996AJ....111.1748F, Wang2018MNRAS.474.1873W, Liu2022ApJ...927...57L}, which does not rely on NELs and thus allows us to extend the search for CLAGNs beyond this redshift limitation.
As described in Sect.~\ref{Recalibration of LAMOST spectra}, we first performed an initial flux scaling using SDSS \texttt{psfMag} measurements to select candidate CLAGNs. For the refined sample, we further utilized quasi-simultaneous photometric data from ZTF, CRTS, and ASAS-SN to improve the spectral calibration and ensure consistency with the contemporaneous flux level. Based on this procedure, we identified a total of 45 CLAGNs, 40 of which are newly discovered in this work. The redshift distribution of our sample spans $0.036 \lesssim z \lesssim 0.338$, with no detections at $z > 0.9$. Despite our calibration method not relying on NELs, the absence of higher-redshift sources likely reflects the limited depth of LAMOST, as high-redshift AGNs generally have lower signal-to-noise ratios, making it difficult to identify changes in BELs.

In terms of spectral transitions, the sample is dominated by turn-off events, with 43 sources exhibiting this behaviour, whereas only 2 sources show turn-on transitions. Similar results have also been reported in previous studies. For example, \citet{Zeltyn2024ApJ...966...85Z} found 76 dimming and 37 brightening CLAGNs in the first-year SDSS-V sample. \citet{Amrutha2024MNRAS.535.2322A} estimated lower-limit CLAGN rates of 1.7\% for turn-on events and 9.6\% for turn-off events. Similarly, only 8 of the 51 CLAGNs reported by \citet{Dong2025ApJ...986..160D} were turn-on events.
This asymmetry is physically understandable. A decline in accretion activity can reduce the ionizing continuum and weaken the BELs, causing a Type 1 AGN to undergo a turn-off transition. By contrast, a Type 2 AGN will likely not undergo a turn-on event because it can either lack a detectable BLR or have its BELs obscured by circumnuclear dust. In such cases, even if the accretion activity increases, observable BELs may not emerge.

\subsection{The incidence of repeating changing-look behaviour}

We cross-matched these 40 newly identified CLAGNs with DESI DR1 spectroscopy and found that 14 sources have three-epoch spectral coverage (SDSS, LAMOST, and DESI). Of these, 7 objects exhibit clear repeating CL behaviour, corresponding to a high fraction of $\sim50\%$. This unexpectedly high occurrence rate motivates us to further examine whether such behaviour is also present in previously reported CLAGN samples.
We therefore cross-matched the 51 CLAGNs from \citet{Dong2025ApJ...986..160D} with DESI DR1 and identified 16 sources with available DESI spectra, 5 of which are confirmed as RCLAGNs, yielding a comparable fraction of $\sim31\%$. The consistency between these independent samples indicates that repeating CL behaviour is likely common for CLAGNs when multi-epoch spectroscopy is available.

In the general AGN population, the detection rate of CLAGNs is known to be low. For example, \cite{Zeltyn2024ApJ...966...85Z} found that only $\sim0.4\%$ of $\sim29000$ AGNs with repeated SDSS-V spectroscopy exhibit CL phenomena, while \cite{Wang2024ApJ...966..128W} reported that $\sim0.3\%$ of AGNs undergo CL transitions over timescales of 5–20 yr. \cite{Amrutha2024MNRAS.535.2322A} further derived turn-on and turn-off CLAGN detection rates of $\sim1.7\%$ and $\sim9.6\%$, respectively, which should be regarded as lower limits due to incomplete temporal sampling that may have missed transitions occurring and reverting between spectroscopic epochs. At higher redshifts ($z > 0.9$), \cite{Chen2026ApJS..282...28C} reported that the fraction of CL quasars is as low as $\sim0.042\%$ based on DESI DR1 and SDSS DR18 spectroscopy. Although these rates vary depending on sample selection and methodology, they consistently indicate that CLAGNs are rare in the overall AGN population. In contrast, our results suggest that the fraction of repeating CL behaviour among known CLAGNs is higher. This high detection rate of repeating CL behaviour suggests that CL transitions are likely driven by recurrent physical processes, such as accretion rate fluctuations or disk instabilities. It also implies that CL phenomena are more common in AGNs than currently observed, with their low overall detection rate limited by observational cadence and time baseline.

\subsection{Timescale}

Table \ref{Table2} summarizes the spectroscopic observation epochs of each CLAGN in the SDSS, LAMOST, and DESI surveys. Since most sources have only two or three spectroscopic observations, the current sample provides limited constraints on the intrinsic timescale of CL transitions. The time interval between two spectra can only be treated as an upper limit on the transition timescale, rather than a direct measurement of the underlying evolutionary process.

\begin{figure}
\centering
\includegraphics[width=0.4\textwidth]{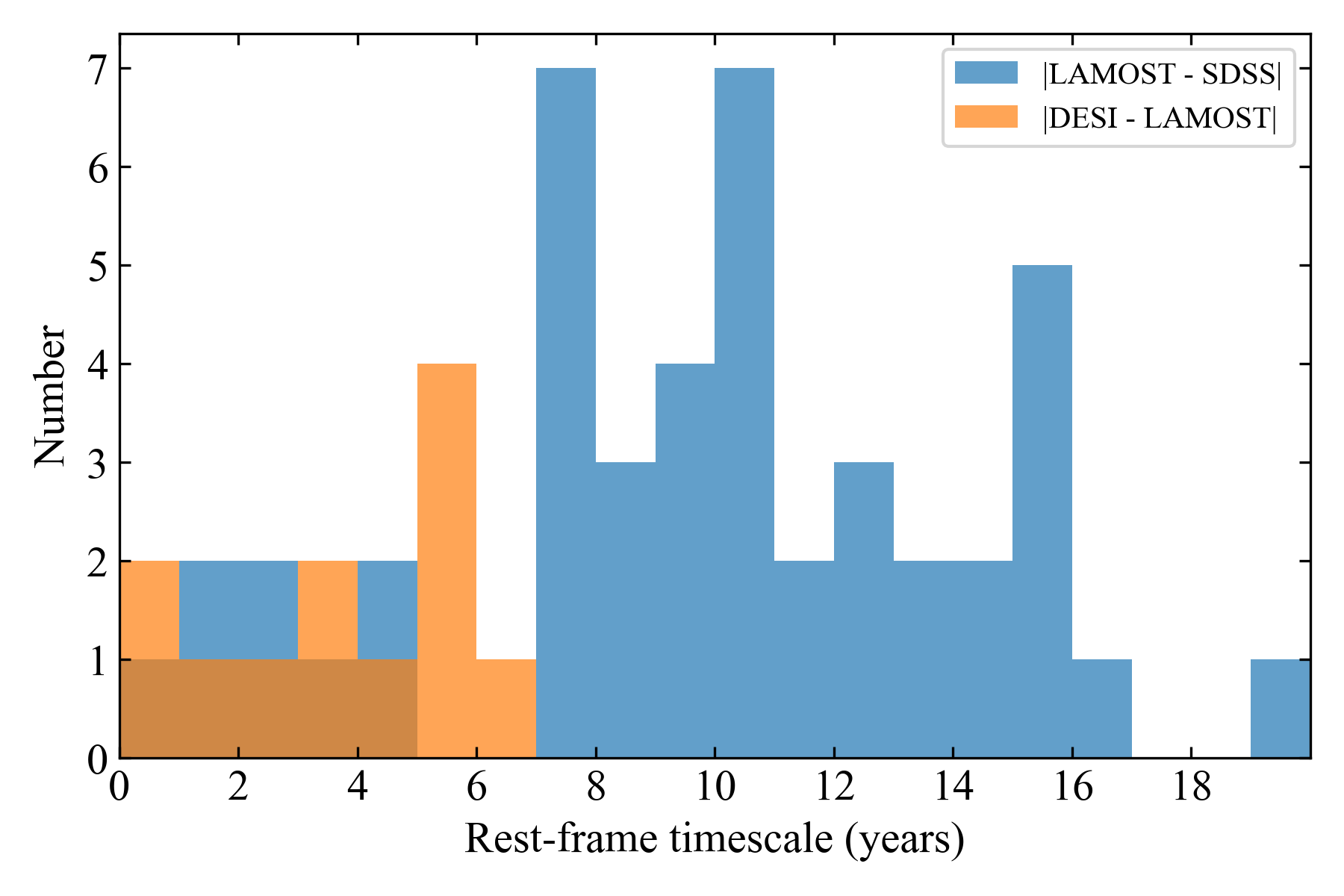}
\caption{Rest-frame distributions of the upper-limit transition timescales for CLAGNs and RCLAGNs.
The blue histogram represents the first spectral-state transition for the 45 CLAGNs, calculated as
$|{\rm MJD}_{\rm LAMOST}-{\rm MJD}_{\rm SDSS}|/[365.25(1+z)]$.
The orange histogram shows the second transition for the 12 RCLAGNs, calculated as
$|{\rm MJD}_{\rm DESI}-{\rm MJD}_{\rm LAMOST}|/[365.25(1+z)]$.}
    \label{fig.4}
\end{figure}

Figure \ref{fig.4} presents the rest-frame distributions of the upper-limit timescales for the first and second spectral-state transitions. The orange histogram represents the first-transition timescales for the 45 CLAGNs, while the blue histogram shows the second-transition timescales for the 14 RCLAGNs.
The rest-frame upper-limit timescales of the first transition are clustered around $\sim$10 yr, while those of the second transition are shorter, with a characteristic value of $\sim$4 yr.

This difference reflects the different time baselines of the spectroscopic survey pairings. The SDSS DR17 spectroscopic data used in this work span approximately from 2000 to 2021 \citep{Abdurro2022ApJS..259...35A}, providing the earliest observational epochs in our sample. The LAMOST DR12 data extend from 2011 to 2024, and therefore partially overlap with SDSS while also reaching later epochs. As a result, SDSS--LAMOST pairings provide long temporal separations, leading to longer inferred transition timescales for the first transition ($\sim$10 yr). In contrast, the second transition is constrained by LAMOST--DESI pairings. DESI DR1 includes spectroscopic observations obtained from 2020 to 2022 \citep{DESI2022AJ....164..207D}, which overlap with the later phase of the LAMOST. This stronger temporal overlap results in shorter observational baselines, leading to shorter transition timescales for the second transition ($\sim$4 yr).

Similar distributions of CLAGN transition timescales have been reported in previous studies \citep[e.g.][]{Yang2018ApJ...862..109Y, Guo2024ApJS..270...26G, Wang2024ApJ...966..128W, Dong2025ApJ...986..160D, Chen2026ApJS..282...28C}, which further support the interpretation that the observed timescale distributions are shaped by survey window functions and heterogeneous temporal baselines, rather than reflecting intrinsic differences in the physical timescales of accretion-state transitions.

\subsection{Black hole mass and Eddington ratio}

Under the assumption that the BLR gas is gravitationally bound to the supermassive black hole, the virial theorem gives a simple estimate of the black hole mass:

\begin{equation}
M_{\rm BH} = f \frac{R_{\rm BLR} \Delta V^2}{G},
\end{equation}

\noindent where $f$ is a dimensionless scaling factor, $R_{\rm BLR}$ is the BLR radius, $\Delta V$ is the velocity dispersion of the BEL (we adopted the FWHM), and $G$ is the gravitational constant.

Instead of directly measuring $R_{\rm BLR}$, we adopted the empirical single-epoch mass estimator based on the broad H$\alpha$ emission line \citep{Greene2005ApJ...630..122G}:

\begin{equation}
M_{\rm BH} = 2.0 \times 10^{6} 
\left( \frac{L_{\rm H\alpha}}{10^{42}~{\rm erg~s^{-1}}} \right)^{0.55}
\left( \frac{{\rm FWHM}_{\rm H\alpha}}{10^{3}~{\rm km~s^{-1}}} \right)^{2.06}
M_{\odot}.
\end{equation}

\noindent Similarly, for the broad H$\beta$ emission line, we adopted

\begin{equation}
M_{\rm BH} = 3.6 \times 10^{6} 
\left( \frac{L_{\rm H\beta}}{10^{42}~{\rm erg~s^{-1}}} \right)^{0.56}
\left( \frac{{\rm FWHM}_{\rm H\beta}}{10^{3}~{\rm km~s^{-1}}} \right)^{2.0}
M_{\odot}.
\end{equation}

\noindent The emission-line luminosities were derived from the integrated line fluxes using
\begin{equation}
L_{\rm line} = 4\pi d_L^2 F_{\rm line},
\end{equation}
where $d_{L}=(1+z)\cdot\frac{c}{H_{0}}\cdot\int_{1}^{1+z}\frac{1}{\sqrt{\Omega _M x^3+1-\Omega _M}}dx$ denotes the luminosity distance, $F_{\rm line}$ is the integrated flux of the corresponding BEL.

\begin{figure*}
\centering
\includegraphics[width=0.7\textwidth]{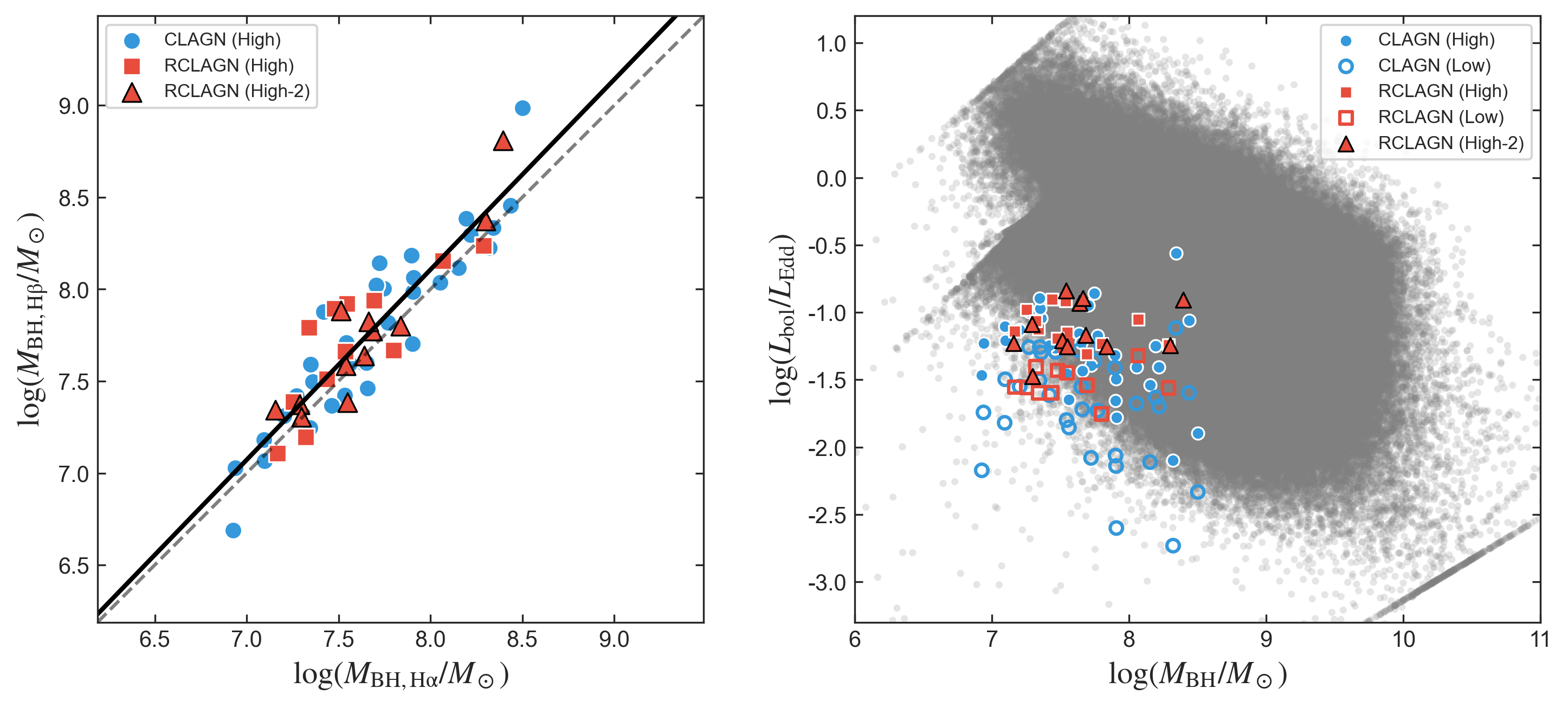}
\caption{\textit{Left}: Comparison of black hole masses derived from H$\alpha$ and H$\beta$ BELs for CLAGNs and RCLAGNs. Blue circles represent single-transition CLAGNs, and red squares and red triangles denote the first and second high states of RCLAGNs, respectively. The dashed black line indicates the one-to-one relation ($y=x$), and the solid black line shows the best linear fit (slope = 1.03, $R^2 = 0.84$).
\textit{Right}: Distribution of CLAGNs and RCLAGNs in the $\log M_{\rm BH}$–$\log (L_{\rm bol}/L_{\rm Edd})$ plane. The grey points represent typical quasars from the SDSS DR16 quasar catalogue \citep{Wu2022ApJS..263...42W}. Blue- and red-filled symbols indicate high-state CLAGNs and RCLAGNs, respectively, while open symbols show their low-state measurements.}
    \label{fig.5}
\end{figure*}

In this work, we used both the H$\alpha$ and H$\beta$ BELs from the high-state spectra to estimate the black hole mass and compare the results. The linear fit shown in the left panel of Fig. \ref{fig.5} yields a slope of 1.03 and an intercept of $-0.04$, with $R^2 = 0.84$, indicating a strong correlation and overall consistency between the two estimates. Since the H$\alpha$ is stronger than H$\beta$ and provides more reliable measurements of the integrated flux and line width, we adopted the black hole masses derived from H$\alpha$ for subsequent analysis.

For the RCLAGNs, the virial black hole mass estimates derived from the first and second high-state spectra are consistent within the measurement uncertainties, despite the substantial changes in broad-line luminosity associated with the intervening CL transitions. Since the true black hole mass is not expected to change over these timescales, this consistency suggests that the single-epoch mass estimates are not strongly affected by the CL state in our high-state spectra. A similar result was recently reported by \citet{Carpenter2026MNRAS.548ag607C}, who found consistent H$\alpha$-based single-epoch black hole mass estimates across three spectroscopic epochs of the CLAGN ZTF18abuamgo, despite pronounced broad-line variability. In a broader AGN sample, however, \citet{Amrutha2026NatCo..17.2385A} found that long-term luminosity variations can lead to changes in single-epoch black hole mass estimates because the broad-line widths do not always respond as expected from BLR breathing. They also found that using the less variable [O III] luminosity to estimate the BLR size improves the consistency of the mass estimates across epochs. These studies indicate that single-epoch virial masses can remain stable across CL transitions in some objects, as found here and by \citet{Carpenter2026MNRAS.548ag607C}, but that long-term luminosity variability remains a potential source of systematic uncertainty. The agreement between the two high-state estimates supports the reliability of our adopted black hole masses, although single-epoch virial estimates still have intrinsic uncertainties.

The Eddington ratio, which serves as an important indicator of the accretion efficiency in the CL phenomenon, is defined as

\begin{equation}
\lambda_{\rm Edd} = \frac{L_{\rm bol}}{L_{\rm Edd}}.
\end{equation}

\noindent Here, $L_{\rm bol}$ is the bolometric luminosity and $L_{\rm Edd}$ is the Eddington luminosity. The latter is given by

\begin{equation}
L_{\rm Edd} = 1.26 \times 10^{38}
\left( \frac{M_{\rm BH}}{M_{\odot}} \right)
{\rm erg~s^{-1}},
\end{equation}
which represents the maximum luminosity at which radiation pressure balances gravitational attraction.

The bolometric luminosity is not measured directly, but is instead inferred from the H$\alpha$ luminosity using empirical scaling relations. In the dim state of CLAGNs, the relative contribution of the host galaxy to the optical continuum becomes significant, making a direct measurement of the intrinsic AGN continuum luminosity at 5100~\AA\ uncertain. We therefore inferred the 5100~\AA\ monochromatic continuum luminosity from the broad H$\alpha$ luminosity using an empirical scaling relation. Specifically, we first adopted the relation between the H$\alpha$ luminosity and the monochromatic continuum luminosity at 5100~\AA\ from \cite{Greene2005ApJ...630..122G}:

\begin{equation}
\lambda L_{\lambda}(5100\text{\AA}) = 2.4 \times 10^{43}
\left( \frac{L_{\rm H\alpha}}{10^{42}~{\rm erg~s^{-1}}} \right)^{0.86}
{\rm erg~s^{-1}}.
\end{equation}

\noindent We then converted the monochromatic luminosity into the bolometric luminosity using a bolometric correction factor:
$L_{\rm bol} = BC_{5100} \cdot \lambda L_{\lambda}(5100\text{\AA})$, where we adopted $BC_{5100} = 9$ following \cite{Kaspi2000ApJ...533..631K}.

The right panel of Fig. \ref{fig.5} shows the distribution of CLAGNs and RCLAGNs in the $\log M_{\rm BH}$–$\log (L_{\rm bol}/L_{\rm Edd})$ plane. Compared with the typical quasars from the SDSS DR16 sample \citep{Wu2022ApJS..263...42W}, both CLAGNs and RCLAGNs are located towards the lower-left region, indicating lower Eddington ratios. In particular, RCLAGNs show an evolutionary trend from high to low and back to high Eddington ratios, suggesting that they experience multiple crossings of the critical accretion regime during their CL phases.

For our sample, the Eddington ratios in the high state span from 0.008 to 0.276, with a mean value of $\sim$0.07, while those in the low state range from 0.002 to 0.077, with a mean of $\sim$0.03. This demonstrates a clear decrease in the accretion rate from the high to low state, consistent with the CL phenomenon being associated with significant variations in accretion power. 
Our results are consistent with previous studies \citep[e.g.][]{MacLeod2019ApJ...874....8M, Green2022ApJ...933..180G, Lyu2022ApJ...927..227L, Wang2024ApJ...966..128W}, which suggest that CLAGNs tend to have lower Eddington ratios than normal quasars and occupy an intermediate regime between luminous QSOs and low-luminosity AGNs.
Interestingly, both the high- and low-state Eddington ratios in our sample are close to the critical value of $\sim 10^{-2}$, at which a transition in the accretion mode has been proposed \citep{Ruan2019ApJ...883...76R}. This critical threshold is analogous to state transitions observed in X-ray binaries, where the accretion flow changes from a radiatively efficient thin disk to a radiatively inefficient mode. The proximity of our sources to this regime supports the scenario that the CL phenomenon is driven by changes in the accretion state.

\section{Summary}
\label{Summary}

In this work we systematically searched for CLAGNs by cross-matching multi-epoch spectroscopic observations from SDSS DR17 and LAMOST DR12. To avoid the redshift limitation imposed by narrow-line-based flux calibration, we first recalibrated the LAMOST spectra using SDSS $gri$ photometry and then refined the flux scaling of visually selected candidates with quasi-simultaneous multi-epoch photometric data. We further incorporated DESI DR1 spectroscopy to investigate repeating CL behaviour. The main results are summarized as follows:

1. We identify a sample of 45 CLAGNs from the SDSS--LAMOST matched spectra, 40 of which are newly reported in this work. The sample is dominated by turn-off events: 43 sources exhibit the disappearance of BELs, while only 2 sources show turn-on behaviour. This asymmetry is physically understandable since Type 2 AGNs can either lack a detectable BLR or have their BEL obscured by circumnuclear dust, making the emergence of observable BELs difficult to identify.

2. By incorporating DESI DR1 spectra as a third spectroscopic epoch, we searched for RCLAGNs. Of the 40 newly identified CLAGNs, 14 have spectra from SDSS, LAMOST, and DESI; 7 of these 14, i.e. $\sim$50\%, show clear repeated transitions. We also cross-matched the 51 CLAGNs reported by \citet{Dong2025ApJ...986..160D} with DESI DR1 and identified 16 sources with available DESI spectra, of which 5 are confirmed as RCLAGNs, i.e. $\sim$31\%. In total, we identify 12 RCLAGNs. This high incidence of recurring CL behaviour suggests that CL transitions are driven by recurrent physical processes, such as accretion rate fluctuations or disk instabilities.

3. The rest-frame upper limits on the transition timescales show different distributions for the first and second transitions. The first transitions, mainly constrained by SDSS--LAMOST, are clustered around $\sim$10 yr, whereas the second transitions, traced by LAMOST--DESI, are concentrated around $\sim$4 yr. This difference is primarily caused by the distinct temporal baselines of the surveys rather than by an intrinsic difference in the physical timescales of the transitions.

4. We estimated black hole masses using the broad H$\alpha$ and H$\beta$ emission lines measured from the high-state spectra. The two mass estimates are in good agreement, with a best-fit slope of 1.03 and $R^2 = 0.84$. Since H$\alpha$ is stronger and more robustly measured, we adopted the H$\alpha$-based estimates as the representative black hole masses of our sample. The resulting black hole masses are distributed over $\log(M_{\rm BH}/M_{\odot}) \sim 6.9-8.5$.

5. The Eddington ratios of our CLAGN sample are lower than those of typical quasars. In the high state, $\lambda_{\rm Edd}$ spans 0.008--0.276, with a mean value of $\sim$0.07, while in the low state it ranges from 0.002 to 0.077, with a mean value of $\sim$0.03. For RCLAGNs, the Eddington ratios evolve from high to low states and then back to high, closely following the repeated spectral transitions. These results support the interpretation that CL behaviour is associated with substantial changes in the accretion power.

\section*{Data availability}
\label{data}
The complete figure set associated with Fig.~\ref{fig.3}, containing the corresponding multiwavelength light curves and multi-epoch spectral comparisons for all 45 CLAGNs, is available on Zenodo at \url{https://doi.org/10.5281/zenodo.21827419}.

\begin{acknowledgements}
This work was supported in part by the National Natural Science Foundation of China (NSFC; grant Nos. 12433004, 12133004, 12203034, and 12303019), the National Key Research and Development Program of China (grant No. 2025YFA1614102), the Eighteenth Regular Meeting Exchange Project of the Scientific and Technological Cooperation Committee between the People’s Republic of China and the Republic of Bulgaria (Series No. 1802), and the China Manned Space Project (grant No. CMS-CSST-2025-A07). 
We also acknowledge support from the Astrophysics Key Disciplines of Guangdong Province and Guangzhou City, the Guangdong Major Project of Basic and Applied Basic Research (grant No. 2024A1515013169), and the Key Laboratory for Astronomical Observation and Technology of Guangzhou.
This work was also supported by the Guangzhou University Graduate Student Innovation Capacity Cultivation Project under Grant No. JCCX2025010, and by the LEPL Shota Rustaveli National Science Foundation of Georgia under Grant No. FR-25-21041.
G. H. Chen (No. 202509940003) and W. X. Yang (No. 202309940006) gratefully acknowledge financial support from the China Scholarship Council. H. B. Xiao acknowledges the support from the Shanghai Science and Technology Fund (22YF1431500) and the Shanghai Municipal Education Commission regarding artificial intelligence empowered research.
\end{acknowledgements}

\bibliographystyle{bibtex/aa} 
\bibliography{references} 

\end{document}